\documentclass[psfig,10pt]{article}

\usepackage{fullpage}
\usepackage{DejaVuSansMono}
\usepackage[T1]{fontenc}

\usepackage{hyperref}
\usepackage{amsmath}
\usepackage{graphicx}
\usepackage{listings}
\usepackage{upquote}
\usepackage{underscore}
\usepackage{microtype}

\usepackage{color}
\definecolor{grey}{rgb}{0.5,0.5,0.5}

\title{Using Memory-Protection to Simplify Zero-copy Operations}
\author{Russell Power}
\date{}

\begin{document}
\maketitle

\begin{abstract} High performance networks (e.g. Infiniband) rely on
  \emph{zero-copy} operations for performance.  Zero-copy operations, as the
  name implies, avoid copying buffers for sending and receiving data.
  Instead, hardware devices directly read and write to application specified
  areas of memory.  Since modern high-performance networks can send and receive
  at nearly the same speed as the memory bus inside machines, zero-copy
  operations are necessary to achieve peak performance for many applications.

  Unfortunately, programming with zero-copy APIs \emph{is a giant pain}.
  Users must carefully avoid using buffers that may be accessed by a device.
  Typically this either results in spaghetti code (where every access to 
  a buffer is checked before usage), or blocking operations (which pretty
  much defeat the whole point of zero-copy).

  We show that by abusing memory protection hardware, we can offer the best
  of both worlds: a simple zero-copy mechanism which allows for non-blocking
  send and receives while protecting against incorrect accesses.
\end{abstract}

\noindent{}To make things concrete, consider an MPI computation that
as part of it's operation maintains a distributed table.  Each worker might have
code like this:

\begin{lstlisting}
// map from int to a big blob
hash_map<int, Buffer[1024]> table;
PutRequest put;
GetRequest get;

while (1) {
  if (world.Iprobe(ANY_SOURCE, PUT_REQUEST)) {
    world.Recv(peer, PUT_REQUEST, &put, sizeof(put));
    memcpy(table[put.key], put.value);
  }
  if (world.Iprobe(ANY_SOURCE, GET_REQUEST)) {
    world.Recv(ANY_SOURCE, GET_REQUEST, &get, sizeof(get));
    world.Send(peer, GET_RESPONSE,  table[get.key]);
  }
}
\end{lstlisting}

(Naturally each worker will probably be doing something else in addition to
maintaining their table.)  This is fairly straightforward, but inefficient code;
the blocking send operation prevents the worker from doing anything else until
the current get request completes.

We'd like to switch to use the non-blocking (ISend) primitive for effiency. A
first attempt at this might look like:

\begin{lstlisting}
list<Request> pending;
while (1) {
...
  if (world.Iprobe(ANY_SOURCE, GET_REQUEST)) {
    world.Recv(ANY_SOURCE, GET_REQUEST, &get, sizeof(get));
    pending.push(world.ISend(peer, GET_RESPONSE,  table[get.key]));
  }
  
  // remove any finished sends
  check_for_completed(&pending);
\end{lstlisting}

Of course, after running this (or if we're smart, before running it), we realize
that there is a possible conflict between our put and get requests.  We can now
handle multiple simultaneous sends, but what if we get a put request for a key
$k$ that we're in the middle of sending?  If this is the only place where table
is used, then we can address this by keeping track of the \emph{active keys}:

\begin{lstlisting}
hash_map<int, bool> active;
while (1) {
  if (world.Iprobe(ANY_SOURCE, PUT_REQUEST)) {
    world.Recv(peer, PUT_REQUEST, &put, sizeof(put));
    while (active[put.key]) { check_for_completed(&pending, &active); }
    memcpy(table[put.key], put.value);
  }
  if (world.Iprobe(ANY_SOURCE, GET_REQUEST)) {
    world.Recv(ANY_SOURCE, GET_REQUEST, &get, sizeof(get));
    active[get.key] = 1;
    pending.push(world.ISend(peer, GET_RESPONSE,  table[get.key]));
  }
  
  // remove any finished sends + clear pending bits
  check_for_completed(&pending, &active);
\end{lstlisting}

What if our table is used elsewhere in our program?  Then we are forced to
either guard every access to the table with a check on our pending table which
leads to ugly, hard to maintain code.  If we have multiple threads reading the
table this becomes even worse; requiring a lock around any accesses, which can
severely compromise performance.

It would be nice if we could somehow ``protect'' any table cells that are still
being sent.  Writes to protected cells would block as appropriate until sends
are finished: 

\begin{lstlisting}
...
 if (world.Iprobe(ANY_SOURCE, GET_REQUEST)) {
   world.Recv(ANY_SOURCE, GET_REQUEST, &get, sizeof(get));
   protect(table[get.key]);
   pending.push(world.ISend(peer, GET_RESPONSE,  table[get.key]));
 }
 // check for finished requests and unprotect
 check_for_completed(&pending, &table);
\end{lstlisting}

We can get part of the way there by utilizing the memory protection hardware
present on our system (exposed via the \textbf{mprotect} system call):
\begin{lstlisting}
void protect(buffer) {
  // we have to align to page boundaries for mprotect
  page_start = align(buffer.ptr)
  prot_len = buffer.size + (buffer.ptr - page_start)
  mprotect(page_start, prot_len, PROT_READ);
}
\end{lstlisting}

This marks a buffer as read-only.  Any writes to it will result in a protection
fault.  On it's own, this is not very useful (except perhaps as a debugging tool
to find errant writes).  We can do better by registering a signal handler for
protection (segmentation) faults.  Our handler is given the memory address that
caused the fault; we just need a mapping of address ranges that are in use:

\begin{lstlisting}
struct Op {
  MPI::Request req;
  void *start;
  int len;
};

hash_map<void*, Op> active;
void handler(int signal, siginfo_t* info, void* ctx) {
  Op& op = get_op(active, info->si_addr);
  while (!op.req.Test()) {
    yield();
  }
  unprotect(op);
}
\end{lstlisting}

We can wrap up sending and memory protection for convenience:

\begin{lstlisting}
void send_and_protect(int dst, int tag, Buffer b) {
  // we have to align to page boundaries for mprotect
  page_start = align(buffer.ptr)
  prot_len = buffer.size + (buffer.ptr - page_start)
  Op op = { world.Isend(dst, tag, b), page_start, prot_len };
  active[page_start] = op;
  mprotect(page_start, prot_len, PROT_READ);
}
\end{lstlisting}

That's it!  Now whenever a writer tries to touch a protected area they will
simply block until the request is finished.  Our code for handling put requests
reverts to the original form:
\begin{lstlisting}
...
if (world.Iprobe(ANY_SOURCE, PUT_REQUEST)) {
  world.Recv(peer, PUT_REQUEST, &put, sizeof(put));
  memcpy(table[put.key], put.value);
}
\end{lstlisting}

The locking of the active regions is now implicit.  Even better, we've also
protected any other areas of our program which might try to write to our table. 

Note that our memory protection boundary is conservative.  Hardware works on
pages (typically 4kbytes) of memory at a time; by aligning our buffers to page
boundaries, we end up blocking operations that don't necessarily conflict with
our sends.  This is not a correctness issue, but it will result in some writes
being blocked unneccesarily.

There are some obvious improvements we can apply at this point. We don't want to
use this technique if our messages are one-byte long: not only will the overhead be far
too high, our false-positive rate will also skyrocket.  A sensible strategy
would be copy messages less then a certain threshold and use protection only on
larger messages.

\section*{Conclusion}

Example code, including an implementation of the above technique and performance
tests is available online here: \url{http://github.com/rjpower/zero-copy}

\section*{Related Work}

There are, of course, many uses (and abuses) for paging and memory protection;
too many to list them all here.  Treadmarks~\cite{amza1996treadmarks} uses the
same protection trick to alias distributed shared memory onto the existing host
address space.  Hardware memory protection has been used as a basis for building
transactional memory systems~\cite{baugh2008using}. Various malloc debuggers use
memory protection to identify reads and writes to freed memory.

\bibliography{ref}
\bibliographystyle{abbrv}
\end{document}